# Thermally stimulated depolarization in hafnium oxide: experiment and numerical simulation

Yu. N. Novikov,[1] D. E. Temnov,[2] E. A. Volgina,[2] M. S. Lebedev,[3] and V. A. Gritsenko[1,4]

*[1]Rzhanov Institute of Semiconductor Physics SB RAS, 13 Lavrentiev aven., 630090, Novosibirsk, Russia*

*[2]Herzen State Pedagogical University, Saint-Petersburg, 191186, Russia*

*[3]Nikolaev Institute of Inorganic Chemistry SB RAS, 3 Lavrentiev aven., 630090, Novosibirsk, Russia*

*[4]Novosibirsk State Technical University, 20 Marx aven., 630073, Novosibirsk, Russia*

This paper presents an experimental and theoretical study of thermally stimulated depolarization (TSD) in hafnium oxide. Three different models for the probability of trap ionization are considered: the Frenkel effect and two variants of the multiphonon trap ionization mechanism, multiphonon ionization mechanism of isolated traps at low trap concentrations and phonon-assisted tunneling between neighboring traps at high trap concentrations. The best agreement between the TSD experiment and theory is observed for the multiphonon ionization mechanism of isolated traps. Similar thermal energies of electron and hole traps in $HfO_2$ are obtained ($W_t^e \approx W_t^h \approx 1.3$ eV).

## 1. Introduction

Hafnium oxide ($HfO_2$) is of not only fundamental [1,2], but also practical [3,4] interest. $HfO_2$ is a key gate dielectric in silicon devices [3,4]. $HfO_2$ is intensively studied for its use in a new generation of high-speed flash memory as a storage medium [5]. $HfO_2$ is used as an active medium in a memory based on the memristor effect in artificial intelligence devices [6].

The overwhelming majority of experiments on the charge transport in dielectric films are interpreted in terms of the ionization of charged Coulomb traps in an electric field (Frenkel model) [7,8]. The Frenkel model, in particular, was used in the interpretation of the current-voltage characteristics of $HfO_2$ [9, 10]. Informative methods of traps spectroscopy in dielectrics are luminescence spectroscopy [11] and thermally stimulated depolarization (TSD) [12]. Experiments on the thermally stimulated depolarization are interpreted on the basis of the Frenkel effect [12]. Recently, the charge transfer in a number of dielectric films [13-15] has been explained by an alternative mechanism – multiphonon ionization of isolated traps (Macram-Ebeid and Lannoo model (ME-L)) [16]. Previously, the ME-L trap ionization mechanism was used to interpret the TSD spectra of $Si_3N_4$ [17]. At high trap concentrations, the distance between traps is small. It is more favorable for an electron to tunnel to a neighboring trap rather than into a conduction band. This multiphonon trap ionization mechanism, at high trap concentrations, is called phonon-assisted tunneling between neighboring traps, or the Nasyrov-Gritsenko (N-G) model [18]. Establishing a reliable mechanism for the ionization of traps in $HfO_2$ is important both for modeling leakage currents in modern field-effect transistors and for predicting the storage time of information in flash memory devices in which $HfO_2$ is used as an active storage medium.

When studying the charge transport in metal-insulator-semiconductor (MIS) structures under the steady-state current, there is an uncertainty regarding the contribution of electrons and holes to the dielectric conductivity. Generally, in an MIS-structure, electrons, which are transported within the dielectric via electron traps, are injected from a negatively biased electrode into the dielectric. At the same time, holes, which are transported within the dielectric via hole traps, are injected from a positively biased electrode into the dielectric. The contribution of electron and hole traps to the steady-state conductivity of the dielectric is uncertain.

The aim of this work is to experimentally and theoretically study the mechanisms of ionization of electron and hole traps in $HfO_2$ using the TSD method. The probability of trap ionization in TSD is analyzed within various physical models.

## 2. Experimental and calculation methods

Thin $HfO_2$ films were deposited on Si (100) substrates (*n*-/*p*-type, resistivities 6–9 and 9.6–14.4 Ω·cm) in an R-200 Advanced ALD reactor (Picosun). Substrates were cleaned by RCA-1 ($NH_4OH:H_2O_2:H_2O$ = 1:1:7) and 10% HF, leaving an ~1 nm native oxide. ALD was performed at 250°C and ~1–2 hPa using tetrakis (diethylamino) hafnium (IV) (TDEAH, 99%, DALCHEM,

Russia) as a hafnium precursor, and $H_2O$ vapor. TDEAH was kept in a bubbler at 90 °C, delivery lines at 100 °C; the $H_2O$ source at 20 °C. $N_2$ (99.999%) served as carrier/purge gas (250 sccm for TDEAH, 100 sccm for others). The pulse sequence was as follows: TDEAH (1.0 s) – purge (10 s) – $H_2O$ (0.5 s) – purge (10 s); 600 cycles were performed. The film thickness (46 nm) was measured by multiangle null-type ellipsometry (LEF-3M ellipsometer, λ = 632.8 nm). The detailed process description, along with the composition and structural characterization, is reported in our recent work [19]. XPS confirmed carbon-free films; XRD/SEM revealed polycrystalline monoclinic grains of ~40-50 nm.

In this study, the dielectric polarization and accumulation of electrons or holes in traps in the dielectric were achieved in a silicon-dielectric-corona plasma system. Corona plasma does not inject electrons or holes into the dielectric [20]. In a silicon-dielectric-corona plasma configuration, electrons or holes are injected from silicon into the dielectric. The electric field in the dielectric is generated by the charged ion deposition on the dielectric surface. A TSC-II system from Setaram (France) was used to measure thermally stimulated depolarization (TSD) currents. The samples were exposed to the electric field generated by a positive corona discharge at a polarization temperature of $T_p$ = 20 °C. Subsequently, the samples were heated at a rate of 9 °C/min to a temperature of $T_m$ = 220 °C. Depolarization currents were recorded using a Keithley 6517A electrometer. All measurements were performed in a vacuum chamber under a helium gas atmosphere. During the TSD measurement process, a small depolarizing electric field was applied to the sample.

To describe the TSD currents in $HfO_2$, transport equations, which are numerically solved together with the Poisson equation [17], are used. In the case where the electric field of the accumulated charge, as a result of sample polarization, is much smaller than the depolarization field, when the depolarization rate is constant, when only one type of carrier (electrons or holes) is involved in the transfer and when a repeated recapture at traps can be neglected, it is possible to use the approximate method developed in [12]. In [12], the formula for calculating the TSD currents, which will be further used to calculate the TSD spectra, was obtained:

$$J(T) = eLn_t P \exp\left( -(\beta)^{-1} \int_{T_0}^{T} P dT \right), \qquad (1)$$

where $J(T)$ is the TSD current density, $T$ is the temperature, $P$ is the trap ionization probability, $T_0$ is the initial sample temperature, $L$ is the sample thickness, $n_t$ is the concentration of electrons captured in the trap in the polarization mode (for holes, the designation $p_t$ is used), $\beta$ is the temperature change rate, $e$ is the electron charge.

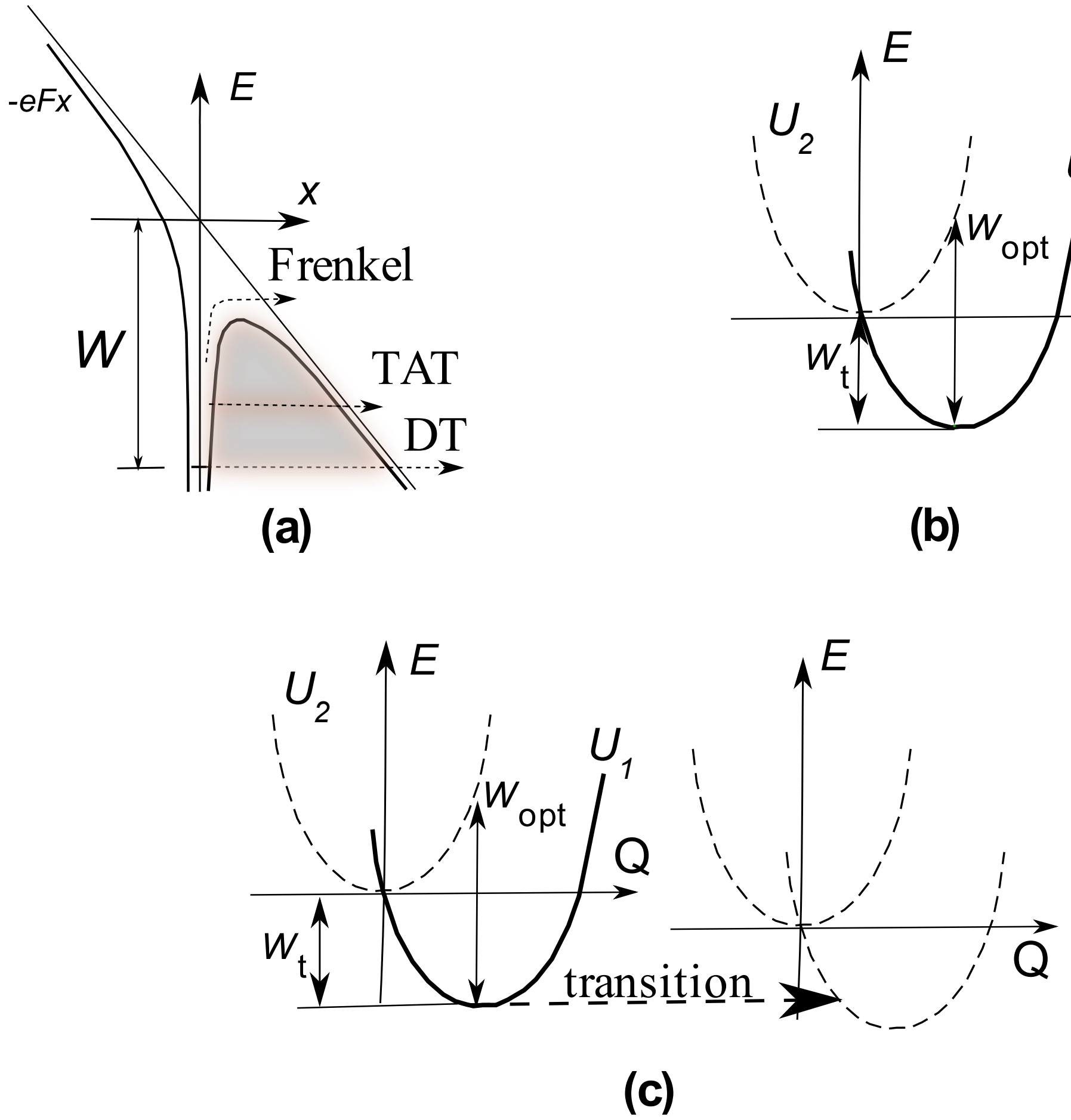

Fig. 1. Energy diagrams illustrating three trap ionization models: a) Frenkel model; here the following notations are introduced: Frenkel effect, TAT — thermally assisted tunneling, DT — direct tunneling, $W$ — Coulomb trap energy; $F$ — electric field; b) multiphonon ionization of isolated traps, low trap concentration, ME-L model; $U_1$ — potential energy of a trap with an electron, $U_2$ — potential energy of an empty trap, $Q$ — configuration coordinate, $W_t$ and $W_{opt}$ — thermal and optical ionization energies of a trap, respectively, c) phonon-assisted tunneling between neighboring traps, N-G model. The dashed arrow (transition) shows the position of oscillators during the electron transition from the ground state of a filled trap to the excited state of an unfilled trap.

In this work, the calculation of the trap ionization probability in $HfO_2$ was performed using three models (Fig. 1): (a) Frenkel model, taking into account thermally assisted tunneling (TAT), (b) ME-L model, (c) N-G model.

In the Frenkel model, the probability of thermal ionization of a trap is determined as in [7, 8]:

$$P_{Frenkel} = \nu \exp\left( -\frac{W-\beta\sqrt{F}}{kT} \right); \quad \beta = \sqrt{\frac{e^3}{\pi\varepsilon_\infty\varepsilon}} \ . \qquad (2)$$

Here, $W$ is the trap energy, $\beta$ is the Frenkel constant, $F=U_{e.h}/L$ is the electric field, $U_e$ and $U_h$ are the depolarization voltage values for $HfO_2$ pre-charged with electrons and holes, respectively, $k$ is the Boltzmann constant, $T$ is the temperature, $\varepsilon_\infty$ = 4.0 is the high-frequency dielectric permittivity [9, 10], $\varepsilon$ is the vacuum permittivity, $e$ is the electron charge, $\nu$ is the frequency factor (Fig. 1a).

In addition to the thermal ionization of a trap through the top of the Coulomb barrier, considered in the original Frenkel's work, we considered the mechanism of thermally assisted tunneling, TAT [15], which, in addition to direct tunneling, provides the tunneling of an electron from the excited state of Fig. 1:

$$P_{TAT} = \frac{\nu}{kT} \int_0^{W-\beta\sqrt{F}} dE \exp\left( -\frac{E}{kT} - \frac{2}{\hbar} \int_{x_1}^{x_2} dx \sqrt{m^*(eV(x)-E)} \right) \ , \qquad (3)$$

$$V(x) = W - \frac{e}{4\pi\varepsilon_\infty\varepsilon x} - Fx \ .$$

Here, $V(x)$ is the Coulomb barrier, $x$ is the coordinate, $m^* = 1.0\, m_0$ is the effective mass of electrons and holes in the monoclinic phase of $HfO_2$ [21], $\hbar$ is Planck constant. The classical turning points $x_1$, $x_2$ were calculated using the formula:

$$x_{1,2} = \frac{1}{2}\frac{W-E}{eF}\left( 1 \pm \left( \frac{eF}{\pi\varepsilon_\infty\varepsilon(W-E)^2} \right)^{1/2} \right) \ . \qquad (4)$$

The ionization rate by the Frenkel mechanism, taking into account TAT, was calculated using the formula [15]:

$$P = P_{Frenkel} + P_{TAT}. \qquad (5)$$

Within the model of multiphonon ionization of isolated traps at their low concentration, the ME-L model (Fig. 1b), the trap ionization probability is given by the expression [16]:

$$P = \sum_{n=-\infty}^{+\infty} \exp\left[\frac{nW_{ph}}{2kT} - S\coth\frac{W_{ph}}{2kT}\right] I_n\left(\frac{S}{\sinh(W_{ph}/2kT)}\right) P_n(W_n), \tag{6}$$

$$P_n(W_n) = \frac{eF}{2\sqrt{2m^* W_n}} \exp\left(-\frac{4}{3}\frac{\sqrt{2m^*}}{\hbar eF} W_n^{3/2}\right), \quad S = \frac{W_{opt} - W_t}{W_{ph}} \ ,$$

where $W_t$ is the thermal and $W_{opt}$ is the optical ionization energy of the trap, respectively, $W_{ph}$ = 0.03 eV is the phonon energy in $HfO_2$ [22], $I_n$ is the Bessel function.

The multiphonon trap ionization probability at a high trap concentration in the N-G model (Fig. 1c) is described by the expression [18]:

$$P = \frac{2\sqrt{\pi}\hbar W_t}{m^* a^2 \sqrt{2kT\left(W_{opt} - W_t\right)}} \exp\left(-\frac{W_{opt} - W_t}{2kT}\right) \times \exp\left(-\frac{2a\sqrt{2m^* W_t}}{\hbar}\right) \sinh\left(\frac{eFa}{2kT}\right), \tag{7}$$

where the parameter $\alpha = N_t^{-1/3}$ corresponds to the distance between the traps for a given concentration of neutral traps ($N_t$).

Both the ME-L and N-G models are based on the multiphonon trap ionization mechanism. The main difference between the N-G and ME-L models is that, in the N-G model, an electron tunnels to a neighboring electron trap, while, in the ME-L model, the electron tunnels the conduction band. The ME-L model is realized at low trap concentrations, while the N-G model is realized at high trap concentrations.

# 3. Experimental and calculation results, discussion

## 3.1. Frenkel model

The experimental dependences of the TSD in $HfO_2$ (squares) with pre-injected a) electrons, $U_e$ = -3.5 V b) holes, $U_h$ = 3.5 V, are shown in Fig. 2.

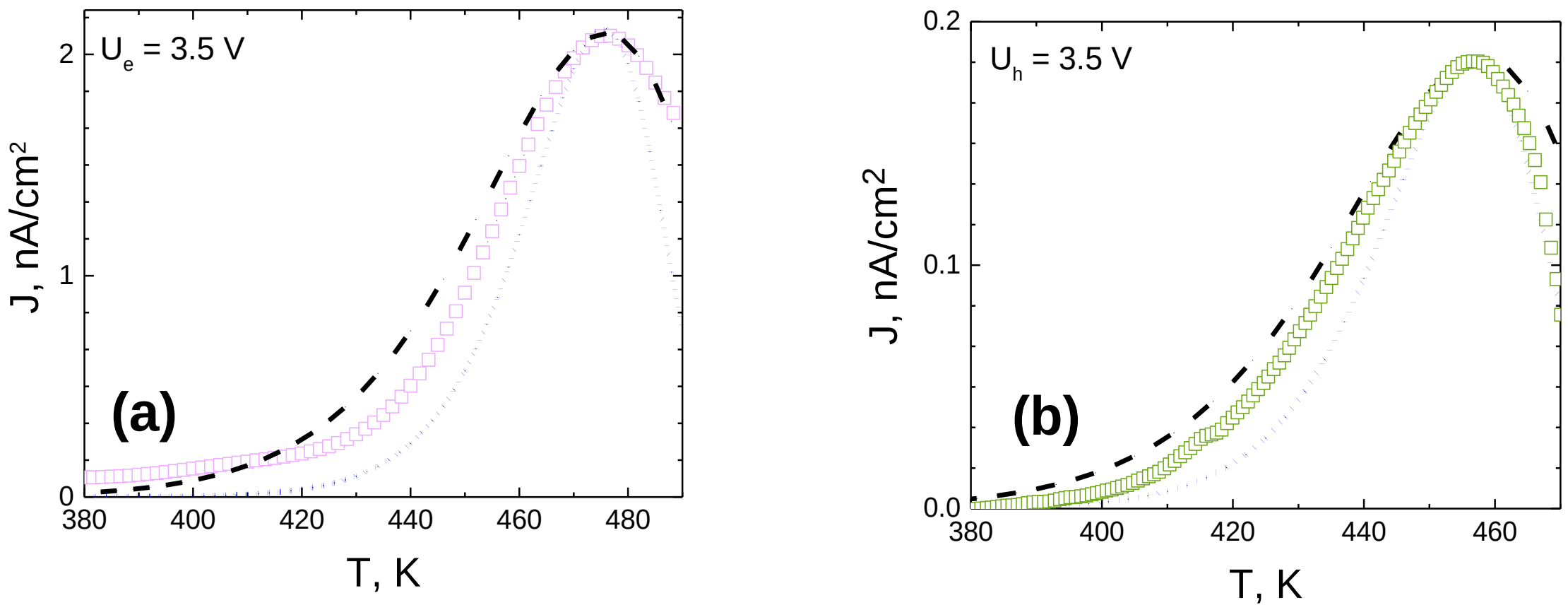

Fig. 2. TSD experiment (squares) and calculation (dashed line and dots) on the Frenkel model taking into account TAT. Calculation parameters: a) the sample was pre-charged with electrons, $U_e$ = -3.5 V; dashed line: $W^e$ = 1.25 eV, $\nu$ = 9 x $10^7$ $s^{-1}$, $n_t$ = 9.4 x $10^{16}$ $cm^{-3}$, dots: $W^e$ = 1.9 eV, $\nu$ = 4.6 x $10^{14}$ $s^{-1}$, $n_t$ = 6.7 x $10^{17}$ $cm^{-3}$; b) the sample was pre-charged with holes, $U_h$ = 3.5 V; dashed line: $W^h$ = 1.25 eV, $\nu$ = 9 x $10^7$ $s^{-1}$, $p_t$ = 9.4 x $10^{16}$ $cm^{-3}$, dots: $W^h$ = 1.84 eV, $\nu$ = 4.45 x $10^{14}$ $s^{-1}$, $p_t$ = 6.1 x $10^{16}$ $cm^{-3}$.

The TSD calculation using the Frenkel model (dashed lines and dots) are shown in Fig. 2. The thermal energies of the electron ($W^e$ = 1.25 eV) and hole traps ($W^h$ = 1.25 eV) used in the calculations are taken from [11]. The calculated TSD peaks in Fig. 2 with traps energy 1.25 eV (dashed lines) are broader than the experimental TSD peaks. In addition, to ensure the agreement between the experiment and theory, it is necessary to use an anomalously small frequency factor value $\nu \sim 10^7$ $s^{-1}$ in the calculations. Previously, within the Frenkel model, when calculating the charge transport in BN [15], $Si_3N_4$ [23] and a number of other dielectrics, anomalously low frequency factor values were obtained. When using the value $\nu = W/h \sim 10^{14}$ $s^{-1}$ in the modeling, the coincidence of the TSD maxima of the calculations with the experiment is reached at a sufficiently high energy values of electron $W^e$ = 1.9 eV and hole $W^h$ = 1. 84 eV traps. Moreover, the width of the calculated TSD spectra (Fig. 2, dots) is smaller than the experimental ones. Calculations using the classical Frenkel model and the Frenkel model, taking into account the TAT, coincide. This is due to a weak depolarizing electric field, under which the contribution of the tunneling component of trap ionization is negligibly small. The anomalously small frequency factor at a thermal trap energy equal to $W$ = 1.25 eV and the small width of the TSD peak at $W^e$ = 1.9 eV and $W^h$ = 1.84 eV, obtained in the calculations, indicate the inapplicability of the Frenkel model for describing the TSD spectra in $HfO_2$.

## 3.2. Multiphonon mechanism of isolated traps ionization at their low concentration, the ME-L model

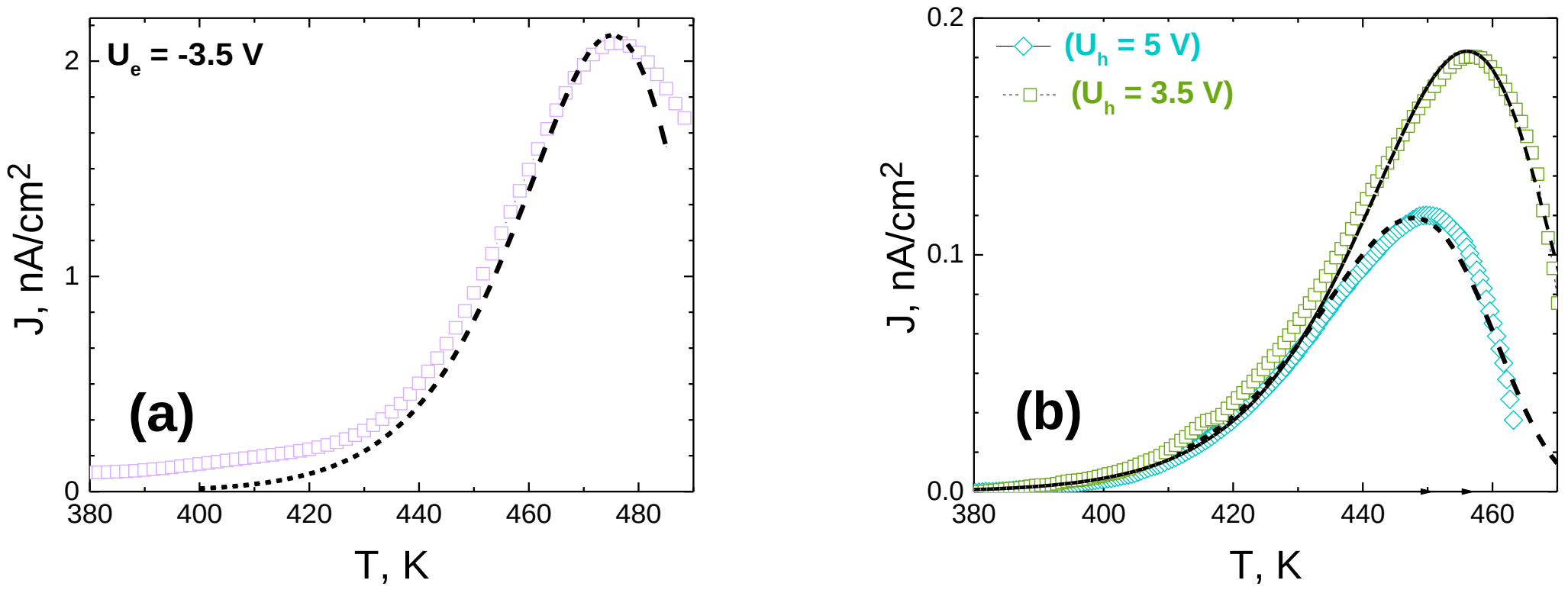

Fig. 3. TSD experiment (squares) and calculation (dashed line) according to the ME-L model. Calculation parameters: a) the sample was pre-charged with electrons: $U_e$ = -3.5 V: $W_t^e$ = 1.39 eV: $W_{opt}^e$= 2.78 eV, $n_t$ = 5.3 x $10^{17}$ $cm^{-3}$; b) the sample was pre-charged with holes: $U_h$ = 3.5 V: $W_t^h$ = 1.29 eV; $W_{opt}^h$= 2.59 eV, $p_t$ = 4.5 x $10^{16}$ $cm^{-3}$ and $U_h$ = 5 V: $W_t^h$ = 1.27 eV; $W_{opt}^h$= 2.54 eV; $p_t$ = 3.0 x $10^{16}$ $cm^{-3}$ .

The experimental dependences of the TSD in $HfO_2$ (squares) with pre-injected: a) electrons, $U_e$ = -3.5 V and b) holes, $U_h$ = 3.5 5 V are shown in Fig. 3. The TSD peak for $U_h$ = 3.5 V is slightly higher than the peak at $U_h$ = 5 V (Fig. 3b) since, in the first case, a slightly larger charge of holes was injected into the $HfO_2$ films. The calculated TSD spectra (Fig. 3, dashed line) are in a satisfactory agreement with the experimental spectra (width, peak position) for reasonable trap parameters: a) the sample was pre-charged with electrons: a) $U_e$ = -3.5 V: $W_t^e$ = 1.39 eV; $W_{opt}^e$=2.78 eV, $n_t$ = 5.3 x $10^{17}$ $cm^{-3}$; b) the sample was pre-charged with holes: $U_h$ = 3.5 V: $W_t^h$ = 1.29 eV; $W_{opt}^h$= 2.59 eV, $p_t$ = 4.5 x $10^{16}$ $cm^{-3}$ and: $U_h$ = 5 V: $W_t^h$ = 1.27 eV; $W_{opt}^h$= 1.54 eV; $p_t$ = 3.0 x $10^{16}$ $cm^{-3}$ . In the literature [21], the values given for the effective mass of electrons vary in the range (0.85 – 1.28 $m_0$) and for holes (1.03 – 1.21 $m_0$). Calculations show that such a "spread" in the effective mass gives deviations from the obtained thermal trap energy value ± 0.005 eV, i.e., a negligible value.

Two voltage values 3.5 and 5.0 V were used to describe the effect of the depolarizing electric field on the TSD spectra for holes (Fig. 3b). An increase in the external electric field leads to an increase in the trap ionization probability [16] and, consequently, to a shift in the TSD maximum toward lower temperatures. From Fig. 3b it is evident that the TSD maximum, at a voltage of 5 V, is shifted toward lower temperatures relative to the TSD spectrum at 3.5 V.

### 3.3. Multiphonon mechanism of traps ionization at their high concentration, the N-G model

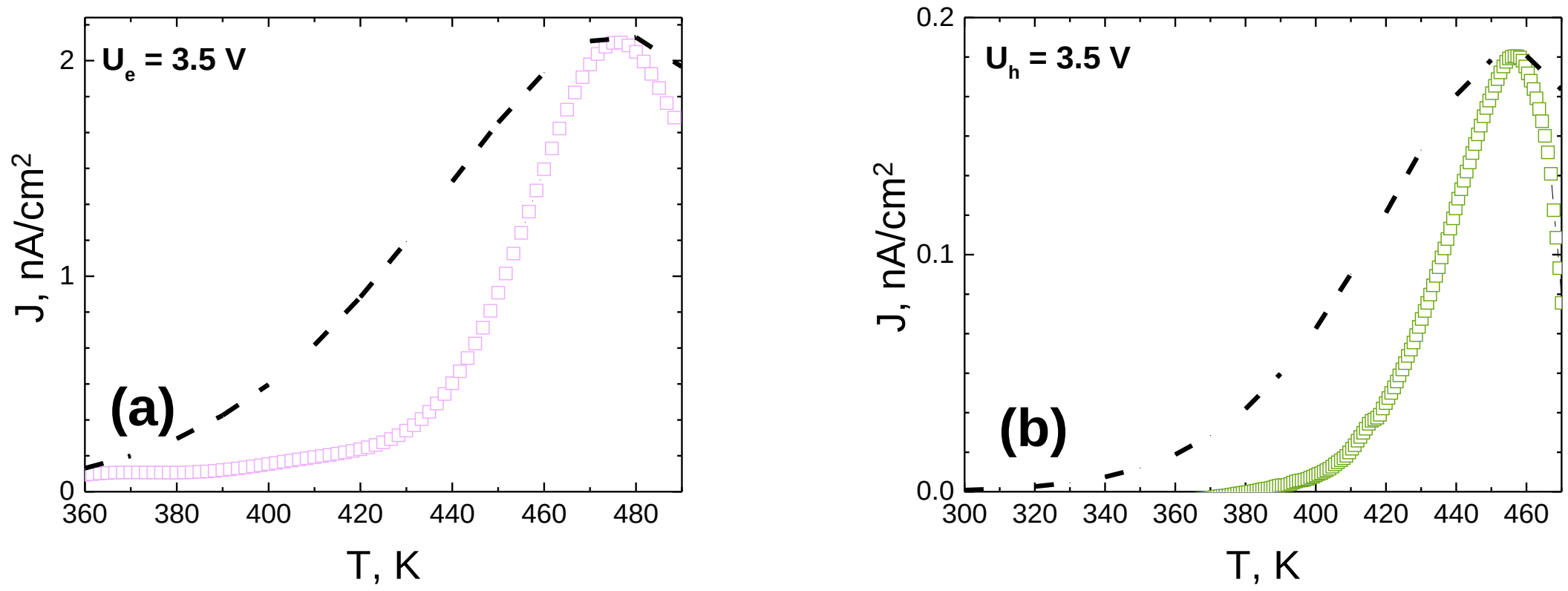

Fig. 4. TSD experiment (squares) and calculation (dashed line) NG model. Calculation parameters: $W_t^e = W_t^h = 1.25$ eV; $W_{opt}^e = W_{opt}^h = 2.5$ eV; a) the sample was pre-charged with electrons: $U_e$ = -3.5 V: $N_t$ = 3.22 x $10^{19}$ $cm^{-3}$; $n_t$ = 2.1 x $10^{18}$ $cm^{-3}$; b) the sample was pre-charged with holes: $U_h$ = 3.5 V: $N_t$ = 3.47 x $10^{19}$ $cm^{-3}$; $p_t$ = 1.7 x $10^{17}$ $cm^{-3}$.

The calculation of the experimental TSD spectra using the NG model is shown in Fig. 4. In formula 7, the trap ionization probability depends exponentially on the thermal energy and on the distance between the traps, i.e., on $N_t^{-1/3}$. A value of 1.25 eV was taken for the thermal energy of electron and hole traps [11]. To ensure that the experimental TSD spectrum coincided with the calculated one, the value of $N_t$ was varied. The calculation of the experimental TSD spectra using the NG model is shown in Fig. 4. The calculation was performed with the following parameters: $W_t^e = W_t^h = 1.25$ eV; $W_{opt}^e = W_{opt}^h = 2.5$ eV; a) the sample was pre-charged with electrons: $U_e$ = -3.5 V: $N_t$ = 3.22 x $10^{19}$ $cm^{-3}$; $n_t$ = 2.1 x $10^{18}$ $cm^{-3}$; b) the sample was pre-charged with holes: $U_h$ = 3.5 V: $N_t$ = 3.47 x $10^{19}$ $cm^{-3}$; $p_t$ = 1.7 x $10^{17}$ $cm^{-3}$. In Fig. 4 it can be seen that the NG model predicts a much broader TSD signal than in the experiment and, therefore, the NG model is not applicable when interpreting the TSD spectra in $HfO_2$.

## 3.4. Results and discussion

The reason for the discrepancy between the calculated TSD curves and the experimental ones (Fig. 2, 4) can be understood using the calculations given in Fig. 5.

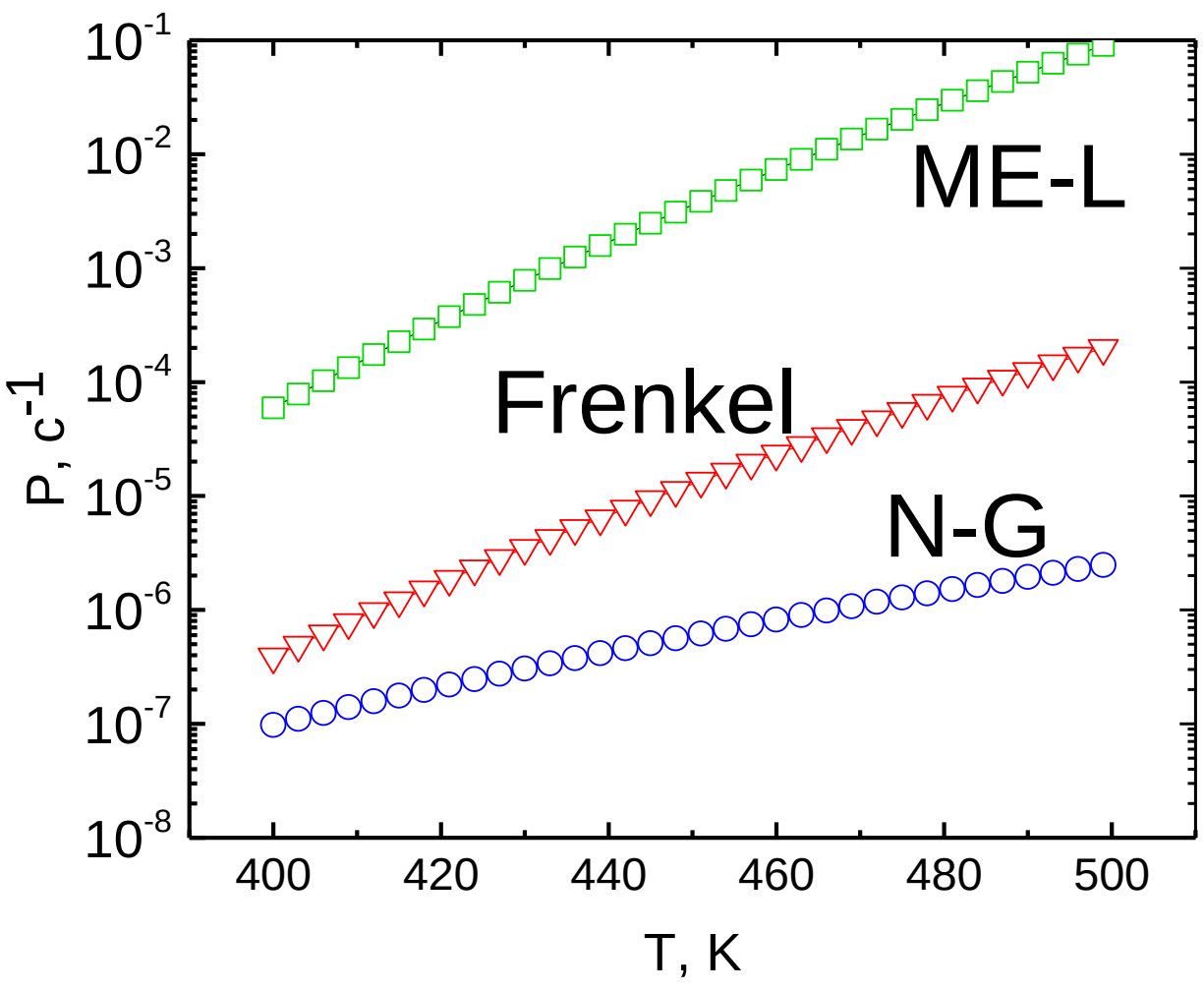

Fig. 5. Comparison of the ionization probabilities of the ME-L (squares), Frenkel (triangles) and N-G (circles) models with the following calculation parameters (taken from Figs. 2, 3, 4): $U$ = 3.5 V: $W_t = W_T$ = 1.25 eV; $W_{opt}$ = 2.5 eV; $W_{ph}$ = 0.03 eV; $N = 3 \times 10^{16}$ cm$^{-3}$; $N_t = 2 \times 10^{19}$ cm$^{-3}$, $v = 9 \times 10^7$ s$^{-1}$.

In Fig. 5 is a comparison of the ionization probability as a function of temperature for different trap ionization mechanisms (Frenkel, ME-L, and N-G) at the depolarizing voltage of 3.5 V. It is evident from the figure that, in the given temperature range and electric field, the ionization probability calculated by the ME-L model is 2-3 orders of magnitude higher than the ionization probability calculated by the Frenkel and N-G models. The low ionization probability in the N-G model is due to the low concentration of neutral traps ($3 \times 10^{19}$ cm$^{-3}$), which corresponds to a large distance between traps. The low ionization probability of the N-G model explains the flat slope of the rise and fall (Fig. 4) of the calculated TSD peak.

Our experimental results on TSD indicate the existence of electron and hole traps with close ionization energies of ≈ 1.3 eV in $HfO_2$. Previously, in [24], the transport of electrons and holes in $HfO_2$ was studied using the injection of minority charge carriers from *n*- and *p*-type silicon substrates. It was found that both electrons and holes contribute to the conductivity of $HfO_2$ [24]. In [11], it is shown that the luminescence band at 2.7 eV and the absorption/excitation band at 5.2 eV are due to an oxygen vacancy in $HfO_2$. The energy of electron and hole traps ($W_t^e = W_t^h$ = 1.25 eV) was determined to be equal to half the Stokes shift of the luminescence spectra [11]. In [25], the electron capture in $HfO_2$ was studied using the pulsed method. A comparison of the experiment and calculations showed that the $HfO_2$ traps energy is 1.3 eV [25]. In [26], using *ab initio* calculations, it was shown that electrons and holes can be captured in amorphous $HfO_2$ in much deeper polaron states than in the crystalline monoclinic phase. Single-electron polarons create bandgap states with energies up to 2 eV. It was shown that deep (more than 1.2 eV) hole traps exist in amorphous $HfO_2$ [26]. In [27], the origin of intrinsic defects present in thin $HfO_2$ films on silicon was experimentally studied using deep level transient spectroscopy (DLTS). It was found [27] that electron traps in $HfO_2$ are oxygen vacancies with energies in the range of 1.22–2.02 eV. The trap energies obtained in [11,25,26,27] are consistent in magnitude with the trap energies obtained using the multiphonon ionization mechanism in this work.

It was shown in [28] that the ratio of optical and thermal ionization energies depends on the defect nature and the strength of trapped electron interaction with phonons. The obtained optical/thermal ionization energy ratio was 2.2 ± 0.3, which is in a good agreement with the value $W_{opt}/W_t$ = 2.0 obtained in this work. This relation between the optical and thermal energy indicates the polaronic nature of traps in $HfO_2$. The polaronic nature of traps indicates that when electrons and holes are captured, the electron energy decreases due to the interaction with phonons, due to the lattice relaxation.

# 4. Conclusions

In this study, various trap ionization probability models were used to describe the experimental TSD currents in $HfO_2$: the Frenkel effect taking into account the TAT, the multiphonon mechanism of isolated traps at low trap concentrations and phonon-assisted tunneling between neighboring traps at high trap concentrations. It was shown that the use of different models significantly affects the TSD peak shape. The Frenkel model does not describe the experimental TSD spectra, since the agreement between the experiment and calculation is reached using an unphysically small frequency factor of $\sim 10^7$ $s^{-1}$. The N-G theory is also unsuitable for calculating the TSD spectra in $HfO_2$ due to the low concentration of neutral traps and, consequently, low ionization probability. The calculations of the TSD spectra based on the theory of multiphonon ionization of isolated ME-L traps are in a satisfactory agreement with the experimental TSD spectra at close energies of ≈ 1.3 eV for electron and hole traps.

**Acknowledgement.** This work was supported by the Russian Science Foundation grant № 25-12-00022.

**Conflict of interest.** The authors declare no conflict of interest.